# Blockchain Enabled Smart Contract Based Applications: Deficiencies with the Software Development Life Cycle Models

Mahdi H. Miraz[1, 2], Maaruf Ali[3]

[1]Centre for Financial Regulation and Economic Development, The Chinese University of Hong Kong, Sha Tin, Hong Kong SAR

[2]School of Applied Science, Computing & Engineering, Wrexham Glyndŵr University, Wrexham, United Kingdom

[3]Department of Computer Engineering, Epoka University, Tirana, Albania

Correspondence: m.miraz@ieee.org

## Abstract

With the recent popularity of Blockchain and other Distributed Ledger Technologies (DLT), blockchain enabled smart contract applications has attracted increased research focus. However, the immutability of the blocks, where the smart contracts are stored, causes conflicts with the traditional Software Development Life Cycle (SDLC) models usually followed by software engineers. This clearly shows the unsuitability of the application of SDLC in designing blockchain enabled smart contract based applications. This research article addresses this current problem by first exploring the six traditional SDLC models, clearly identifying the conflicts in a table with the application of smart contracts and advocates that there is an urgent need to develop new standard model(s) to address the arising issues. The concept of both block immutability and contract is introduced. This is further set in a historical context from legacy smart contracts and blockchain enabled smart contracts extending to the difference between "shallow smart contracts" and "deep smart contracts". To conclude, the traditional SDLC models are unsuitable for blockchain enabled smart contract-based applications.

## Keywords

Blockchain, Smart Contract, Software Development Life Cycle (SDLC), SDLC Models, Distributed Ledger Technologies (DLT).

## Introduction

The concept of smart contract was first put forward in the early 1990s by an American computer scientist and lawyer namely Nick Szabo (Szabo, 1996), obviously without the application of blockchain (BC) technologies. In fact, blockchain was first introduced by Satoshi Nakamoto (Nakamoto, 2008) in the late 2000s as a by-product of the Bitcoin cryptocurrency. This version of the blockchain, also known as Blockchain 1.0 (Miraz, 2019), was implemented without the capability of smart contract. The integration of both was first seen in Blockchain 2.0 (Miraz, 2019). This integration of smart contract with blockchain has exponentially accelerated the use of smart contract and application of blockchain beyond cryptocurrencies (Miraz and Ali, 2018a). Such integration of smart contract retains all the characteristics of the legacy ones and additionally inherits the features of blockchain technologies offering dual benefits. However, the inherited features also include the architectural limitations of blockchain.

The fusion of smart contracts with blockchain assumes blockchain's immutability which provides an extra layer of security. However, immutability of blocks makes it impossible, or at least difficult to some extents, for software engineers to bring any future modification in the application including fixing bugs, future enhancements etc. Therefore, traditional SDLC models do not fit well for BC enabled smart contract-based applications, especially the testing and maintenance cycles. In this paper, the author advocates for developing new standards and models giving attention to blockchain based applications. However, detailed discussion of other design trade-offs (Medellin and Thornton, 2019a; Medellin and Thornton, 2019b) and legal debates (UK Jurisdiction Taskforce 2019; Schmitz and Rule, 2019; Raskin, 2017) on the acceptability of smart contracts as traditional contracts are beyond the scope of this paper.

Contract-based application development and platforms which are both smart and blockchain enabled are uniquely complex and thus significantly differ from those which do not have any connection with the blockchain. This naturally arises because of the following three atypical characteristics of working with the requirements of blockchain based applications, listed below:

i. The immutability of the blocks (of data) being added to the (distributed) ledger;
ii. The internal architecture (software and hardware) for achieving the consensus and conducting the verification and validation i.e. the need for inputs from the peer-to-peer network participants (mining nodes), and
iii. The need for the use of publicly available third-party development platforms, such as Ethereum, which runs on a pay per transaction (trigger of the smart contract) model - making verification and testing very expensive.

Therefore, traditional SDLC models, especially the testing, verification and validation phases of these, do not fit well for this purpose. It is thus pertinent to re-visit the traditional models, identify the mismatch and develop new standards. This paper is principally intended for software engineers as well as blockchain researchers, aiming to partially fill in this gap.

## Methodology

To keep the research focused, only the popular SDLC models were surveyed. This included a detailed investigation with respect to blockchain characteristics of the anatomy of their different phases, especially the testing, validation and verification steps. An impartial examination of the relevant blockchain aspects were conducted avoiding the hype.

To keep the survey focused, keywords such as "SDLV vs. Blockchain", "software development life cycle models", "Blockchain testing and verification" etc. were used in renowned databases such as IEEE Xplore, ACM Digital Library, SpringerLink. These directed keywords were input via both individual and integrated services provided by the Chinese University of Hong Kong E-library. Grey publications such as Masters and PhD theses, blog entries and published materials from unreputable sources were excluded. Only sources from reputable publishers/databases with relevant materials for our research have been included.

This paper explores the discernment between legacy smart contracts and blockchain enabled smart contracts and then investigates how these distinctions makes traditional SDLCs unsuitable to apply for the later. The conceptual framework of this research is best depicted in Fig. 1.

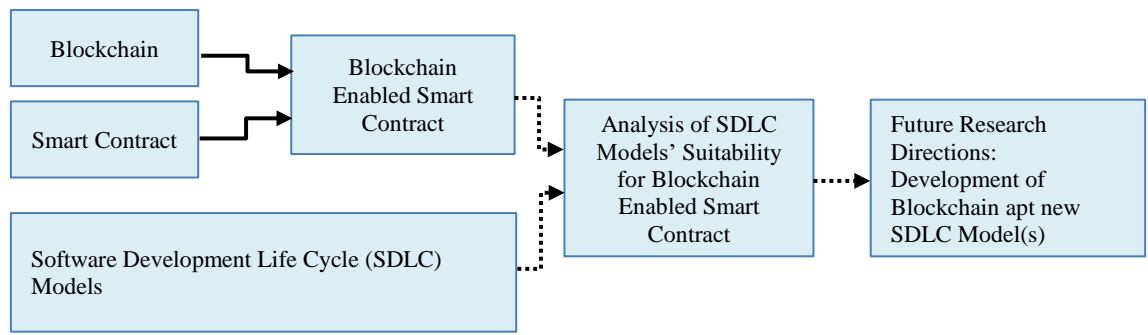

**Figure 1: Conceptual Framework.**

## Smart Contracts

As stated earlier, the term "smart contract" was first coined by Szabo in 1994. He defines smart contracts as contractual clauses embedded in computer systems (hardware, software or both), in a way that makes infringement of a contract expensive (Szabo, 1996). He sees smart contracts as a (computerised) transaction protocol executing the terms of a (legal) contract as per the occurrences of events. The simplest example of a smart contract is a vending machine which displays the prices of the products, accepts payments and dispatches the product (and/or change) based on the users' selections. However, applications of smart contracts now go far beyond the vending machines providing solutions that are more complicated. The 'smart' notion of the term alludes to the capacity to interact with smart systems i.e. computerised protocols.

Blockchain enabled smart contracts are mainly enhanced interpretation of legacy smart contracts with pre-defined conditions but stored in and operates on a blockchain ecosystem, where execution or enforceability of the terms of the contract is automated – without the need of any third-party intermediary for trust.

Based on the degree of automation, smart contracts can be further categorised as "shallow smart contracts" and "deep smart contracts" (Kõlvart, Poola, and Rull, 2016). While shallow smart contracts just perform basic operations such as transaction of a crypto-coin, deep smart contracts are capable of handling more complicated (multiple) operations based on the nature of the trigger(s) or input(s). To summarise, smart contracts are digitally signed computer protocols (contracts) between two or more parties 'representing' a legal contract which can also be based on blockchain or other Distributed Ledger Technologies (DLT). Different schools of scholars define and perceive smart contracts differently – some emphasise on "legal contracts" while others focus on "contract coding". There is no firm or standard definition of smart contracts (Wang, Lau and Mao, 2019).

## Blockchain's Immutability

Immutability of the data blocks is one of the major offerings (Miraz, and Ali, 2018b), amongst all the other benefits offered by the blockchain or distributed ledger technologies. The underlying architecture of the blockchain furnishes immutability as a de facto standard. The mathematical bindings offered by cumulative hashing and digital signature, time-stamp, Proof-of-Work (PoW) or similar other algorithms for determining and providing authorities power to create blocks, consensus mechanism, distributed network – all contributes, significantly if not equally, to the immutability aspect of blockchain data. The following features of a blockchain ecosystem jointly make the ledger immutable (Nakamoto, 2008):

- The header of each block contains the hash of its data segment i.e. of all other data contained in the block. Thus, changing even a single bit in the block data will invalidate the block hash which will invalidate the block itself.
- $n^{th}$ block, in its data segment, directly contains the hash of $(n-1)^{th}$ block, which indirectly is a contribution of the hashes of all the previously sealed blocks in the chain. Thus, changing the contents or data, even by a single bit, of any block will not only invalidate the occurring block but also all the blocks which were created after it.
- Blocks are digitally signed by the private key of the creator (or miner) whose public key is known to the network for verification purposes. Since the private key is only known to that particular node, it is not possible to duplicate the creation of a block by other miners or nodes.
- To validate any transaction or creation of a block, reaching a consensus amongst the participating nodes or miners is required.

- Double spending problem is eliminated by the application of PoW or PoS (Proof of Stake) or other similar algorithms, along with hashing and digital signatures. To be successful to 'double' spend by fooling the system, not only defeating the other honest nodes is enough, but also knowing the private keys of all other nodes or miners as well as reaching a consensus will be needed which is nigh impossible.
- Multiple copies of the chain are stored in the participating nodes. This distributed nature of the system eliminates the possibility of Single Point of Failure (SPF) and makes it inexpedient to hack or deceive the cohort of honest nodes to reach a consensus.

Therefore, blockchain is not just a one-way append only distributed ledger, rather it provides complete immutability by its (technical) architectural design. However, this immutability makes it difficult to bring any changes in any blockchain enabled smart contract-based applications once implemented, such as fixing of any bugs found.

## Software Development Life Cycle (SDLC)

The SDLC models are considered as the foundation software engineering tools to help facilitate better delivery of any software project. SDLC models are predefined set of phases to lay out the common understanding of the complete picture of the development process, from conceiving the concept to final delivery and continued maintenance of software projects or applications. SDLC delineate and pre-establish the way in which the software project will be comprehended and then developed deriving from the business insights and requirement analysis phase to materialise them into features and functions of a software or application – until its operation and application to satisfy and achieve the business needs. Consequently, to assure successful completion of any software project or application, it is very important to select the most appropriate Software Testing life cycles and SDLC model and act in accordance with it, following the requirements and concerns of the project. Amongst the varied SDLC models, with their own strengths and weakness, six of them are more popular, viz.:

  i.  Waterfall Model
 ii.  V-Shaped Model
iii.  Iterative and Incremental Method
 iv.  Spiral Method
  v.  Big Bang Model
 vi.  Agile development.

This article briefly revisits these already established SDLC models and explores their suitability versus contradictions with regards to blockchain enabled smart contracts-based applications.

## Waterfall Model

The Waterfall is the pioneer model in introducing the concept of the SDLC. There are variations or modification of the original model (see Fig. 2). However, the fundamental concept remains the same i.e. dividing the software development tasks into various sequential phases where the progress of the tasks follow from the top to bottom without overlapping amongst each other, similar to a cascading waterfall.

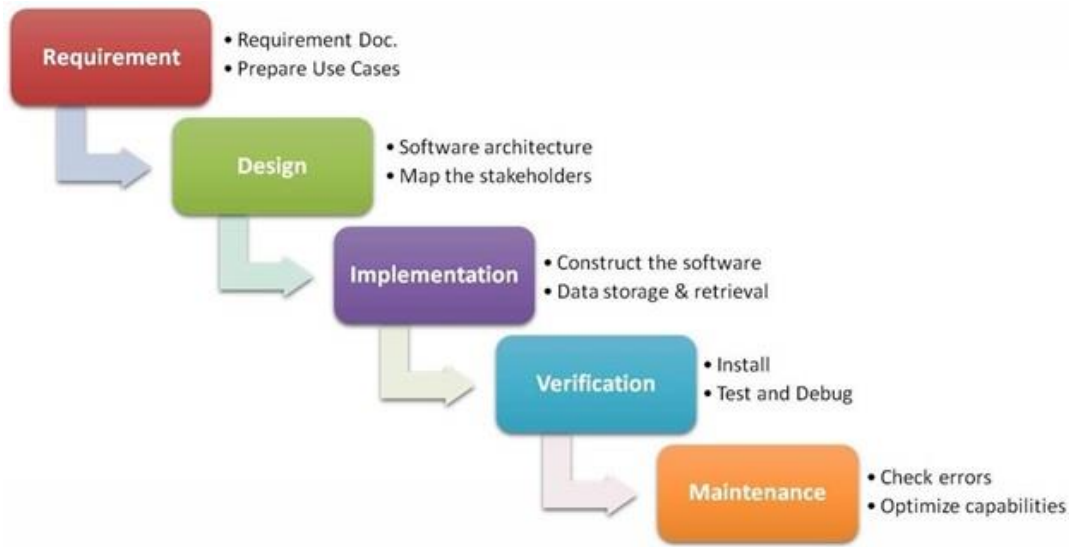

Figure 2: Unmodified Waterfall Model[1].

Detailed discussion of the various steps of the model is out of the scope of the paper. Therefore, to keep the discussion focused, only the relevant steps and analysis of them are given considering the immutability and relevant aspects of blockchain enabled smart contract applications. In the (system) "Design" face of the development model, the codes to be written in the next phase are usually created. The next i.e. "Implementation" phase taking input from the system design phase develops the small units, to be integrated in the next step. Each unit also goes through a testing process, namely Unit Testing, at this stage. At this point, it is very important to consider the fact that the blockchain ecosystem is built on distributed networks, where input from the miners or the participating nodes play a vital role in the verification, validation and consensus reaching processes for both transactions (data) and creation of blocks. Therefore, while some of the units can be tested individually, many of them highly depend on the (collective) inputs from others requiring a complete implementation of the system. Thus, waterfall models fail to comply with the nature of blockchain applications.

---

[1] https://images.ukdissertations.com/118/0518331.001.jpg [Accessed 17 Jan 2020].

"Maintenance" at regular intervals for future enhancement or fixing any bugs or flaws found, is the final phase of this model. Since codes of the smart contracts are stored in the blocks which are immutable, updating them requires modifying the already created blocks, making it impossible due to the immutable nature of blockchain. Various off-chain scaling solutions have been developed and proposed, however, there is no simple mechanism to fully satisfy the needs.

## V-Shaped Model

The V-Shaped Model is mainly an extension of the Waterfall Model. Instead of a complete leaner approach, as adopted in the Waterfall Model, the steps follow an upward curve starting at implementation or coding phase forming the shape of the English letter "V" (see Fig. 3). Another aspect of the extension of the model, from the originating Waterfall Model, is the inclusion of a set of verification and validation steps. The model, as shown in Fig. 3, demonstrates the relationship of the verification phases of the development process with their respective testing (validation) phases (Sami, 2012). The implementation or coding remains in the centre of the curve.

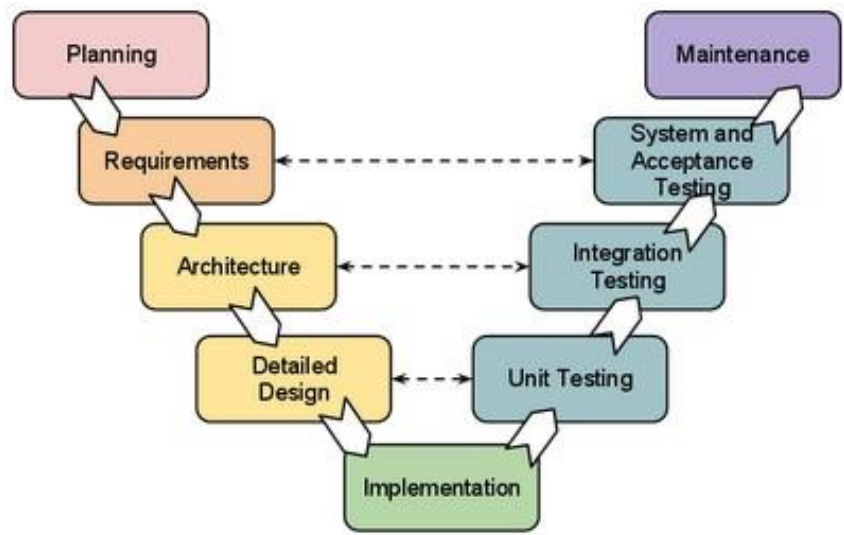

**Figure 3: V-Shaped Model (Sami, 2012).**

As the V-Shaped Model extends the Waterfall Model, it inherits the disadvantages (and also advantages) of its predecessor – those particularly relevant to the development of blockchain enabled smart contract-based applications are the important parameters of discussion of this paper. Apart from those discussed in the Waterfall Model section above, the additional tests at the module or unit labels makes it comparatively more unfit for adoption in such applications. Additionally, the acceptance testing is conducted in the user environment which requires actual implementation of the application.

## Iterative and Incremental Model

Antithetical to Waterfall model consisting of inflexible step-by-step leaner development phases, Iterative and Incremental SDLC Model is a cyclical process blended approach – any combination of iterative design/method with incremental build model, as shown in Fig. 4.

After the completion of the inceptive planning stage, a small handful of steps are recurrently repeated with the aim to incrementally improve the software at each iteration or completion of cycle. However, Iterative and Incremental model requires extensive engagement with the users resulting in increased pressure. In some cases, this is a hapless obligation, as new iterations require testing and feedback from the previous one for evaluating the required changes properly. Therefore, this approach is not a good fit for blockchain enabled smart contract-based applications. In a blockchain ecosystem, users are usually the participating nodes (miners) in the distributed networks. Involving such users at each iteration for validation and verification in each phase is highly impractical. Furthermore, the events of the smart contracts may also be triggered by inanimate objects such as IoT devices and sensors (Ghodoosi 2019) – making the model more unsuitable for this purpose.

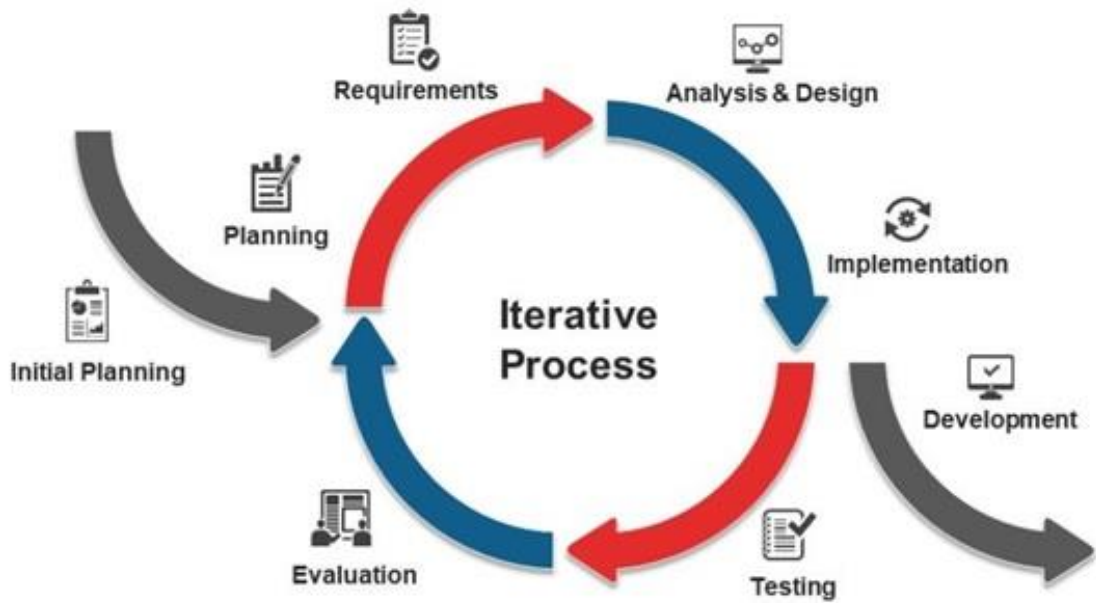

**Figure 4: SDLC Iterative Model[2].**

---

[2] https://www.slideteam.net/media/catalog/product/cache/960x720/i/t/iterative_process_model_Slide01.jpg [Accessed 17 Jan 2020].

# Spiral Model

The spiral model offers a combination of iterative development of the project from evolutionary adoption of prototype model, with the systematic and controlled features of the sequential waterfall model. The spiral model is more suited for large scale projects requiring consistent improvement and refinement over each iteration around the spiral. The output of the specific activities of one iteration is a small proof-of-concept (POC) prototype, a part of the large software, which is used to gather user feedback. In the subsequent spirals, the same activities are repeated, with refinement of the POC prototype, to produce a working model of the software called build having version ID/number. Each versions of the build are sent out to the users for getting feedback for further enhancement in the next version until the final product is ready. Figure 5 demonstrate this as a pictographic representation of the spiral model.

Because user involvement is needed at each iteration, the spiral model suffers from the same suitability problem as discussed for the iterative and incremental model, as discoursed in the section above. Apart from the user involvement, setting up the blockchain based prototype or versioned build at each iteration over the spiral in a distributed network, possibly with IoT devices and sensors as trigger acting devices, is very complicated. Therefore, this feature of the spiral model is not a good fit for blockchain enabled smart contract-based applications.

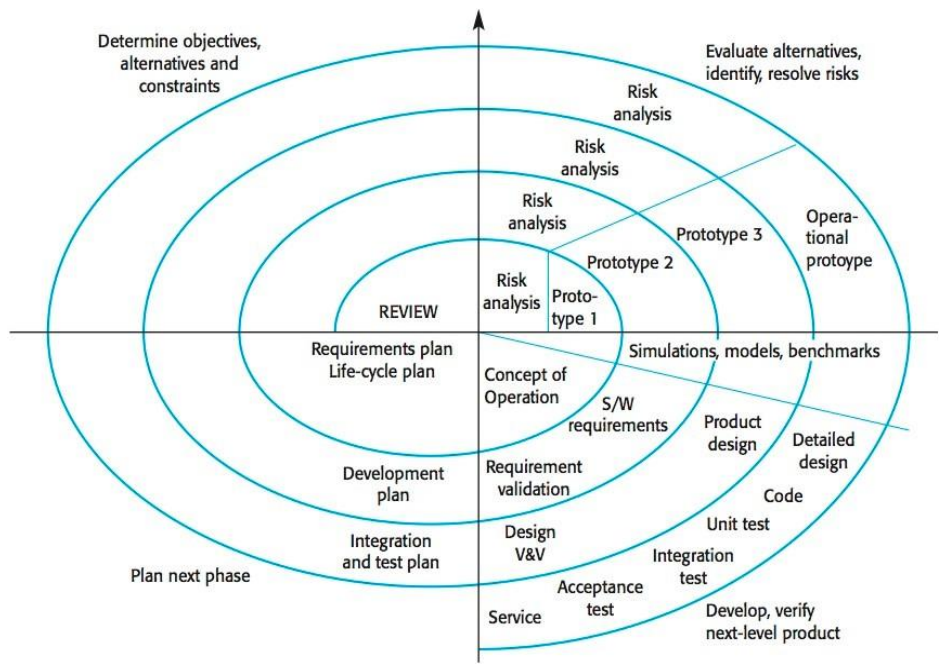

**Figure 5: Spiral Model[3].**

---

[3] https://www.learntek.org/blog/wp-content/uploads/2019/05/spring-phases-1.jpg [Accessed 17 Jan 2020].

## Big Bang Model

Following the notion of the cosmological theory of the Big Bang, software development using Big Bang SDLC model begins with 'nothing' but is followed by emergence of expeditious expansion and growth of code leading to the quick production of a finished product. The Big Bang Model, as shown in Fig. 6, is considered as a unique approach, as unlike other SDLC models, it does not require any firm plan or organisation and does not follow any typical protocol or specific procedure. Rather it follows a "hit the ground" approach – starting the project with immediate effect and quick completion. The major disadvantage of using the Big Bang model is the associated risk, it is highly critiqued for being extremely risky. Any unforeseen issues can lead to sudden complicacy which could have been mitigated easily if built on previous versions or prototypes, or at least if some planned testing and verification were conducted. In the worst-case scenario, serious bugs or errors in the code or system may completely halt the software. Programming for smart contract being relatively new, this poses even higher risks. Various monetary services beyond cryptocurrencies, are being materialised using smart contract based blockchain applications such as initial coin offering (ICO). Hacking of such applications based on Decentralized Autonomous Organizations (DAO), has been a big concern in the recent past (Walch, 2017). Even though the blockchain ecosystem is very secure by its architecture, loopholes in the coding of the smart contract can easily lead to such hacking, therefore, the Big Bang model is not a good fit.

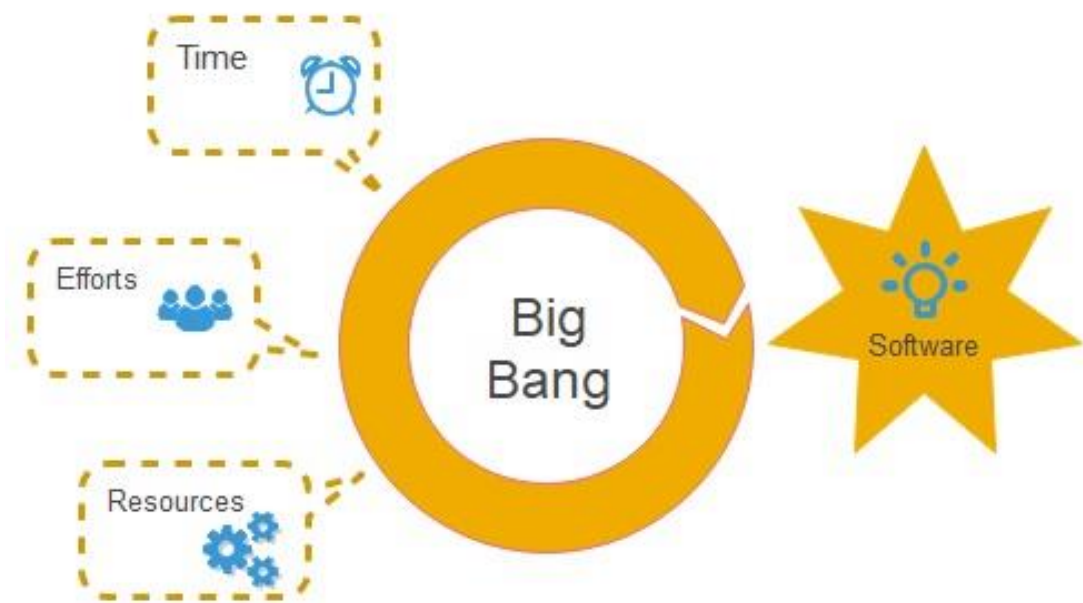

**Figure 6: Big Bang Model[4].**

---

[4] https://static.javatpoint.com/tutorial/software-engineering/images/software-engineering-big-bang-model.png [Accessed 17 Jan 2020].

## Agile Model

To mitigate the 'rigidness' observed in the traditional SDLC model, as discussed in the above sections, and to provide the developers with 'agility', a relatively new approach i.e. the Agile SDLC model has been developed. The agility aspect mainly refers to the ability to quickly adapt to transforming requests and scale-up to the future requirements. Therefore, fast and easy project achievement adopting a flexible approach is the major endeavour of Agile software development.

As shown in Fig. 7, the Agile model combines iterative and incremental software development process models, with the aim to quickly deliver a working model of the software to the customer with agility to future enhancements and modifications.

The Agile SDLC model always welcomes any requirements for new addition of feature or modification of the old ones. Since the codes of the smart contracts are written in the immutable blocks, this feature does not suit blockchain enabled smart contract applications. Furthermore, consensus of the users or the participating nodes may also require bringing in any future change. Agile SDLC models require simultaneous development via cross functional teams including pair programming, extra layer of complicacy is added to the development process. High customer collaboration is required for Agile SDLC model, which as discussed before, is not always practical in a distributed application like blockchain.

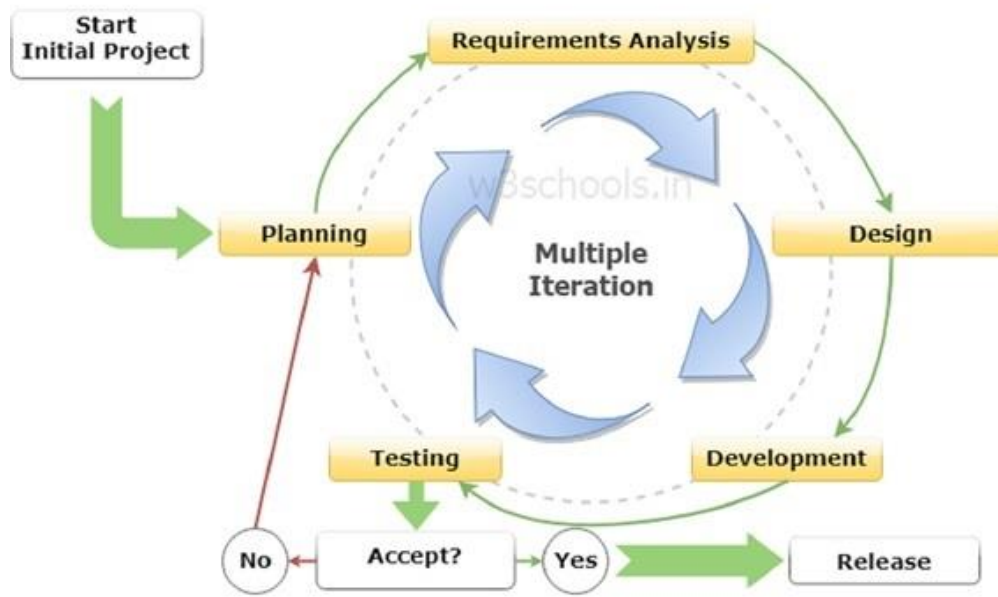

**Figure 7: Agile Development Model[5].**

---
[5] https://www.w3schools.in/sdlc-tutorial/agile-model/

## Results and Discussion

In the previous section six popular SDLC models were briefly introduced and discussed as to why these models fail to completely satisfy the needs of the blockchain enabled smart contract based applications. Table 1, summarises the main propositions of the above discussion and research. Furthermore, all contracts are incomplete (Ayres and Gertner, 1989), including the 'smart' ones. There are two major reasons behind this incompleteness: the contracting parties fail to clearly identify all the future contingencies or the current contract is "insensitive to relevant future contingencies." (Siegel, 2016).

Table 1. Main Problems of the Six Common SDLC Models to Blockchain Enabled Smart Contract based Application Development.

| # | Model | Phase Affected | Comment |
|---|---|---|---|
| 1 | **Waterfall** | System Design | Advance Coding is required – inconvenient due to the overall blockchain architecture. |
|   |   | Implementation | Unit Testing – required user involvement at each unit is inappropriate considering how blockchain functions. |
|   |   | Maintenance | Versioning – immutability of the ledger causes inconvenience. |
| 2 | **V-Shaped Model** | Unit Testing | The required user involvement at each unit is inappropriate considering how blockchain functions. |
|   |   | Acceptance Testing |   |
| 3 | **Iterative and Incremental Model** | Every Phase | User Involvement – extensive user involvement to get feedback is strenuous for blockchain development and applications. |
| 4 | **Spiral Model** | Proof-of-Concept (POC) Prototypes | The distributed nature as well as immutability makes installation and modification very complicated. |
|   |   | Versioned Builds |   |
| 5 | **Big Bang Model** | Overall Process | Associated risks of bugs and thus potential hacking of smart contracts. |
| 6 | **Agile Model** | Agility | In addition to the above issues, reaching consensus may require bringing in changes. |
|   |   | Customer Collaboration | Required user involvement at each unit is inappropriate considering how blockchain functions. |

These were actually proven by the 2016 hack of DAO in which 3.6 million Ether (Ethereum crypto-currency), then equivalent to approximately $50 million US Dollars, were drained into a "Child DAO". This is to note that, the event of the 2016 hacking, or most of the other smart contract hacking thus far, were not due to lack of architectural security offered by blockchain, rather the inherent loopholes in the codes of the smart contract let the hacker manipulate the system. In fact, the hacker later argued that the transfer of funds was completely legal because the smart contracts are self-arbitrating – outsiders does not have the authority to modify the transaction rules. (Siegel, 2016). Thus, the SDLC models which possess high risks such as the Big Bang model, are not suitable for applications containing blockchain enabled smart contracts.

Another point of concern is the complexity of calculating the Gas price (Mulders, M. 2018), the cost needed for running a smart contract on a public blockchain such as the Ethereum platform, especially for large-scale projects with complex coding of smart contracts. Therefore, the SDLC models suitable for large scale projects such as the Agile model, becomes less suitable for such application, especially if there are budget constraints.

The Gas price and other associated costs for smart contract testing needs to be carefully considered while selecting the SDLC model. Because, in most cases, smart contracts are run on a public blockchain platform such as Ethereum, the participating nodes (miners) are incentivised for validating and varying the transactions. The Gas price and similar other prices acts as the source of such incentives. Therefore, each test such as unit testing, acceptance testing etc. at each iteration costs money. As a result, SDLC with high testing frequency will end up costing a fortune. On the contrary, setting up an own blockchain environment is highly expensive for most of the software development firms, thus making it impractical.

In contrast to most of the SDLC models, instead of rigorous testing of the blockchain enabled smart contract-based applications, emphasis is rather given on examination of the codes by professional experts, for practical reasons, as stated above. Considering the fact that, developing smart contact-based applications is an emerging trend, lack of in-house programmer(s) may lead to examination of the coding to be outsourced to external experts. In addition, recent trends in using cloud computing based Blockchain-as-a-Service (BaaS) platforms (Onik and Miraz, 2019), makes it further difficult to conduct in-house tests. Therefore, instead of fully-flexed testing, dependency on examinations of code is highly increasing - leading to higher possibilities of vulnerability.

The blockchain based applications highly suffer from scalability due capped latency of the PoW algorithms and reaching consensus for high volume transactions as well as interoperability amongst chains/ledgers (Donald and Miraz, 2019). This leads to extra complicacy at the testing phases as prescribed in various traditional SDLC models. Recent developments of Atomic Swap and Lightning Network (Miraz and Donald,

2019a; Miraz and Donald, 2019b) techniques holds great potentials in addressing these problems, however, there is still a long way to go before they are fully eliminated.

## Conclusions

In this article, the fundamental differences between blockchain based smart contracts and legacy smart contract applications adopting traditional software engineering approach were studied. The study then examined the suitability of adopting various traditional SDLC models and software testing and verification approaches for the purpose of developing blockchain enabled smart contract-based applications. It has been found that, due to the "immutability" feature of blockchain enabled smart contracts, inherited from the underlying architecture of distributed ledger technologies, these traditional SDLC models are not a good fit for such applications. Therefore, this paper recommends the need for revised SDLC model(s), especially designed to meet the requirements of blockchain based applications.

To conclude, the traditional SDLC models are unsuitable, at least to some extents, for blockchain enabled smart contract based applications due to the immutability it offers, complexity of coding, the distributed nature of the blockchain ecosystems, the consensus required for bringing chances, verification and validation process to be carried by the participating nodes etc. Therefore, the authors of this article advocate the need for developing new SDLC models, particularly tailor-made for blockchain based applications. Collectively the deficiencies in the six SDLC models expressed by other researchers for each model support the results presented in this paper. This is that the developed model(s) then need to be thoroughly evaluated by experts for small to medium to large scale real projects. For obvious reasons, this is out of the scope of this article, however, this clearly shows the future research directions.

## Acknowledgements
The author would like to thank Professor David C. Donald and Professor Stephen Hall, both from the Faculty of Law, The Chinese University of Hong Kong - for their opinions on the legal matters of smart contracts.